\documentclass{jpsj2}
\usepackage{graphicx}

\title{
Making a Universe
}

\author{
S. Horata\thanks{E-mail address : horata\_shinichi@soken.ac.jp} and 
T. Yukawa\thanks{E-mail address : yukawa@soken.ac.jp}
}

\inst{
Hayama Center for Advanced Studies, \\
The Graduate University for Advanced Studies (Sokendai), \\
Hayama, Kanagawa 240-0193, Japan

}

\recdate{\today}

\abst{
For understanding the origin of anisotropies in the cosmic microwave
background, rules to construct a quantized universe is proposed based
on the dynamical triangulation method of the simplicial quantum
gravity. 
A $d$-dimensional universe having the topology $ D^d $ is
created numerically in terms of a simplicial manifold with $d$-simplices
as the building  blocks. 
The space coordinates of a universe are identified on the boundary surface
$ S^{d-1} $, and the time coordinate is defined along the direction 
perpendicular to $ S^{d-1} $.
Numerical simulations are made mainly for 2-dimensional universes, and
analyzed to examine appropriateness of the construction rules by
comparing to analytic results of the matrix model and the Liouville
theory. 
Furthermore, a simulation in 4-dimension is made, and the result
suggests an ability to analyze the observations on anisotropies by
comparing to the scalar curvature correlation of a $ S^2 $-surface formed as
the last scattering surface in the $ S^3 $ universe.
}
\kword{ Origin of Universe, CMB, Inflation, Quantum Gravity, 
Dynamical Triangulation, Numerical Simulation}

\begin{document}
\maketitle

\section{Introduction}
Recent observations of anisotropies in the cosmic microwave background
(CMB) by COBE and WMAP \cite{COBE,WMAP,WMAP-DATA} extend our knowledge on the
history of the universe beyond the big bang. 
They imply as follows: about 14 billion years ago, the universe
started expanding with extremely rapid acceleration, now known as {\it
inflation} \cite{Guth, Sato}, from a micro-universe of quantum
mechanical scale.
Because of the expansion speed exceeding the light velocity, the pattern of original quantum
fluctuations in space could not be disturbed dynamically, and was kept
intact during the expansion something like a picture printed on an
inflating balloon. 
After a little while($ 10^{-33} \sim 10^{-32} $ sec) the expansion was 
slowed down through coupling 
to matter fields, and fluctuations developed in the space curvature
 transferred to matter density fluctuations. 
They eventually are observed as the temperature anisotropies in the CMB,
caused by the photon energy dissipation in fluctuating media known as
the Sachs-Wolfe mechanism \cite{SACHS-WOLFE}. 

Although the observed data are rather well-understood by known
physical processes with several phenomenological parameters, questions
such as how and what kind of fluctuations were created initially, and
why the universe could expand so rapidly are left to be the matters 
which essentially belong to the fundamental lows of space-time. 
In the history of studies on quantum gravity it is the most
remarkable occasion where the theory is required to be
phenomenological, apart from describing quantum fluctuation of
geometry, the {\it graviton}, whose appearance is beyond the current
experimental scope.
It is a common belief that there exists no consistent theory that
combines quantum mechanics and gravity at the moment. 
Among various efforts to establish quantum theory of gravity the
super-string is regarded to be the most promising candidate as far as
quantization of the graviton is concerned. 
It is extraordinary that the superstring theory can unify all the
known interactions on the way of quantizing graviton. 
However, it does not say much about how the space-time is created. 
While absence of a consistent theory of quantum gravity is true in
4-dimension, we certainly have consistent theories of quantum gravity
at least in 2-dimension. 
One of them is a standard field theory called the Liouville field
theory \cite{LIOUVILLE}, and the other is the matrix model
\cite{MATRIX-MODEL}. 
In addition, the latter model has a numerical realization known as the
dynamical triangulation (DT) \cite{DT2D}. 

The existence of analytic theories is closely tied to the
dimensionality of space, and it is not a straightforward task at all
to extend those theories into higher dimensions. 
The conformal gravity \cite{CG,HS} is one of possible generalizations of
the Liouville theory into 4-dimension. 
It is free from space singularities such as the black hole of quantum
mechanical size, and claimed to be renormalizable. 
However, it is unfortunate that the non-unitary nature inherited in
any higher derivative field theory masks charms of the theory\cite{HHY}. 
An extension of the discretized space model into higher dimensions
beyond the matrix model has been proposed for the 3-dimensional space
by adopting the $6j$-symbols as the elementary unit of space
\cite{TURAEV-VIRO}.
In an analogous manner it may be possible to extend further to the
4-dimensional space by introducing an appropriate elementary unit\cite{OGURI}.
The simplicial quantum gravity is one of the constructive
generalizations in the discretized space approach, selecting a
$d$-simplex as the elementary block of the $d$-dimensional space. 
Numerical simulations have been performed for the $ 2, 3,$ and 
$ 4 $-dimensional simplicial spaces\cite{Lattice}, and the grand 
canonical Monte Carlo simulations have produced significant results 
not only for the 2-dimensional
case \cite{GC-DT2D}, but also for 4-dimension \cite{GC-DT4D} in
calculations of the string susceptibility which plays the role of a
universal observable of the space geometry. 
As far as the possibility of extending the method into 4-dimension is
concerned the simplicial quantum gravity is one of the promising cases so far
exhibiting its applicability.

In modern field theories, quantization is often carried out by the
Feynman path integral in which the Euclidean metric is adopted for
definiteness of integrations.
For a system having the stable vacuum it is always possible to
transform the metric from the Euclidean to the Lorentzian and {\it
vice versa}, by picking up one coordinate, which is regarded as the
time coordinate, and making the Wick rotation. 
However, this procedure cannot be applied for a system without the
stable vacuum.
Quantum gravity is one of the systems having non-stationary vacuum of
expanding space. 
Even though there exists no standard method of quantizing such a system,
 we employ the path integral method as a possibility to a space with open
topology.
While our final goal is to establish a method to construct the
quantized universe in the 4-dimensional space-time, the
following discussions are mainly proceeded on two dimensional 
space concerning the difficult circumstance of knowing no guiding theory. 
Besides its simplicity, analytic studies compiled extensively in
2-dimension are inevitable for achieving some hints required in this
totally new subject. 
We would like to stress, however, that the constructive nature of the 
method will make it readily applicable to higher dimensional cases as
well. 

In the next section, we review the DT method mostly concentrating
 on the 2-dimensional space with the $ S^2 $ topology, and show that it
provides a natural extension to a system with the open topology $ D^3 $. 
In Sec.~3, we shall practice to build an ensemble of 2-dimensional
quantum universes based on two fundamental rules. 
In order to check the construction method we calculate observables of
the space for systems with the stable vacua, by tuning the
cosmological constant so that we can compare them to analytical
results of the matrix model and the boundary Liouville theory
\cite{FZZ}. 
In Sec.~4 we define the time and space coordinates of
fluctuating manifold, and calculate the two point correlation function
 for the expanding universe. 
In the last section we give a summary and discussions on future
directions such as a method extending into 4-dimension and
problems related to the introduction of matter contents of the
universe. 
There, we show a numerical result of the simulation in 4-dimension,
which clearly indicates the applicability of the present method to the CMB
anisotropies.
%

\section{Methods of simplicial quantum gravity and their extension}
The dynamical triangulation (DT) has been a standard method of
non-perturbative quantum gravity \cite{dt-gen} known as the simplicial
quantum gravity. 
It was intensively studied in 2-dimension mainly because of its close
connection to the string theory.
Although we cannot expect new results from the 2-dimensional DT more
than those already known from the matrix model and the Liouville
theory, the achieved consistent results encourage us to adopt this
method as a reliable technique for investigations beyond
2-dimension\cite{HHY}. 
One of the drawbacks of the simplicial quantum gravity used to be the
restriction to spaces with closed topologies with the Euclidean
metric. 
In the followings, we would like to show how to extend the
$d$-dimensional DT technique to a space with the open topology 
$ D^d $ for $ d \geq 2 $. 
Besides the space-time topology the dynamics requires a proper
signature of the metric, which is known to be Lorentzian in this
universe. 
We regard the selection of the signature should come out naturally 
when we consider dynamics, since the correct signature, which is required by dynamical laws, supposed to emerge in the process of establishing the space-time. 

Quantum gravity is formally defined in a form of the path integral,
\begin{equation}
Z=\int {\cal D} [g_{ab}] \exp (-S[g_{ab}]). \label{eq2.1}
\end{equation}
Here, we consider the simplest case that the metric tensor $ g_{ab} $
is the unique dynamical variable. 
The space is assumed to have the dimension $ d $ and the 
topology $ S^d $ unless specified.
The simplicial quantum gravity interprets the path integral to be a
sum over all the distinct configurations of space discretized by
 $ d $-simplices. 
For a $ d $-dimensional manifold the action $ S[g_{ab}] $ is given by
a linear combination of the variables, $ \{ N_n \} $ with $ n=0 \cdots d $. 
It is considered to be natural, since we
regard the diffeomorphysm invariance as its independence of the local
geometry, or in other words, local variables.
Thus only global variables such as the total number of $ n $-simplices 
$ N_n $ are proper, and linearity comes from the extensive property 
of the action function.
The $ (d+1) $ variables $ \{N_n\} $ are not all independent. 
The Euler relationship and the manifold conditions reduce the number 
of independent variables.
For example in 2-dimension, they are $ N_0-N_1+N_2 = \chi $ with 
$ \chi $ being the Euler characteristic, and $ 3N_2=2N_1 $.
This makes the action to be $ S^{(2)} = \mu N_2 $ with a parameter$ \mu $, which we call the lattice cosmological constant.

\begin{figure}
\begin{center}
\includegraphics{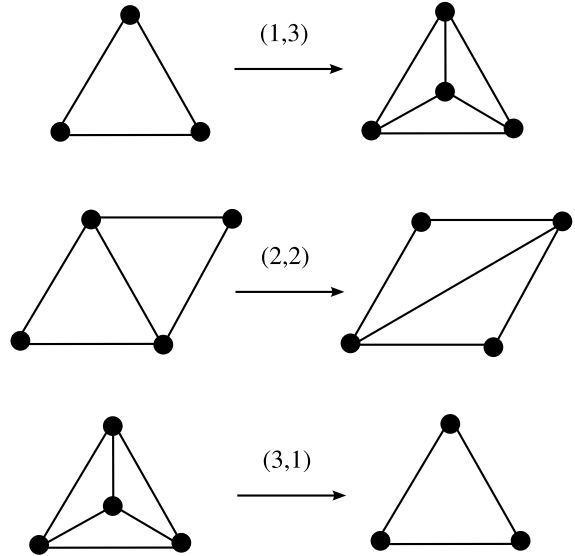}
\caption{$(p,q)$-moves in the 2-dimensional simplicial gravity.}
\end{center}
\label{FIG_PQMOVE}
\end{figure}

The set of distinct triangulations is collected computationally by the
Metropolis Monte Carlo method creating a Markov chain of
configurations $ \{ \alpha_i \} $,
\begin{equation}
\alpha_0 \rightarrow \cdots \rightarrow \alpha_i \rightarrow 
\alpha_{i+1} \rightarrow \cdots. \label{eq2..2}
\end{equation}
Among possible methods of generating configurations one after
another, we employ the so-called $ (p,q) $-moves which allow us a
systematic extension into higher dimensions. 
These moves are known to conserve the topology.
There are three types of $ (p,q) $-moves in 2-dimension as shown in
Fig.~\ref{FIG_PQMOVE}.
From the figures it is clear that $ p $ and $ q $ stand for the
numbers of triangles disappeared and appeared by the $ (p,q) $-move,
respectively. 
%
The Metropolis Monte Carlo method employs the importance sampling,
accepting or non-accepting a new configuration created from an old
configuration by a randomly selected move according to the detailed
balance condition,
\begin{equation}
 \frac{p_{\alpha}} {n_{\alpha}} w_{ \alpha \rightarrow \beta} 
  = \frac{p_{\beta}} {n_{\beta}} w_{\beta \rightarrow \alpha}, \label{eq2.3}
\end{equation}
with the given probability $ p_ {\alpha}  $ for a configuration $ \alpha $. 
The factor $ n_{\alpha} $ counts the number of possible moves starting
from a configuration $ \alpha $. 
This is an important factor for simulations with varying the number of the 
degree of freedom in
order to obtain the proper statistical weight for each configuration.

The number of distinct triangulations is known to grow exponentially in $ N_2 $ for 2-dimension, and the partition function behaves as
\begin{equation}
 Z \sim N_2^{\gamma-3} \exp \left\{ - (\mu-\mu_c) N_2 \right\} \label{eq2....4},
\end{equation}
where $ \gamma $ is historically called the string susceptibility.
As $ \mu $ approaches to a critical value $ \mu_c $ from above, the
average value of $ N_2 $ tends to diverge as $ (\mu-\mu_c)^{-1} $, and
the system reaches to the continuum limit where the simulation meets
the Liouville field theory with the field theoretical cosmological
constant,
\begin{equation}
 \lambda= \frac{\mu-\mu_c}{\frac{\sqrt{3}}{4} a^2},
\end{equation}
for space with a fixed volume $ A= \frac{\sqrt{3}}{4} a^2 N_2 $, where
$ a $ is the lattice constant.
If we choose the parameter $ \mu $ below $ \mu_c $, configurations
with more triangles become favorable, and $ N_2 $ rapidly increases as
a simulation proceeds.
Although it is beyond cases where the standard quantization procedure
is applicable, we regard the partition function obtained by this
method as the quantization of a system with an unstable vacuum,
simply because the method is a trivial extension of quantization for a
system with the stable vacuum.

Another important setup we need to modify for a practical application
of the method to a system with an unstable vacuum is the topological
structure. 
The simplicial quantum gravity usually consider the space manifold
within a closed topology. 
Of course we may regard it corresponds to a case of oscillating
universe, if we take the space topology seriously as a real universe. 
However, it is widely believed that our universe has an open topology
and it is expanding indefinitely in both space and time. 
Then, we wish to extend the DT method so that we can create the space
with a boundary such as the $d$-disk $ D^d $.
In order to show how it is possible, let us take the 
2-dimensional simplicial gravity as an example. 
Among the three $(p,q)$-moves we examine the $ (1,3) $-move, firstly. 
%
%
\begin{figure}
\begin{center}
\includegraphics{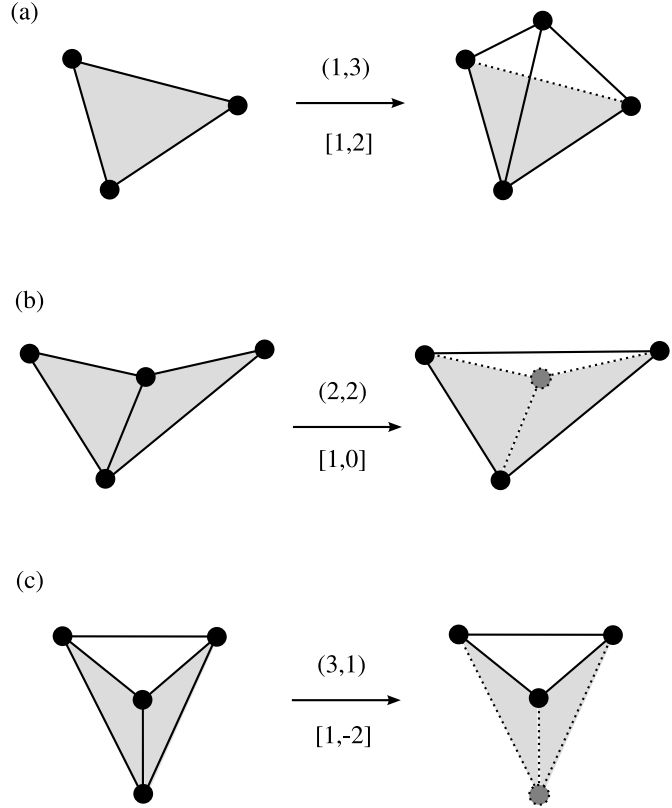}
\end{center}
\caption{$ (p,q) $-moves in $ S^2 $ and $ [\Delta V, \Delta S] $-moves 
in $ D^3 $.}
\label{FIG_S2MOVE}
\end{figure}
%
This move can be described either by dividing a triangle into three
triangles by adding a vertex in the triangle, or by attaching one
tetrahedron on a triangulated surface with one triangle face-to-face,
which we call the $ [1,2] $-moves representing
changes of the volume $ \Delta V $ and the surface area $ \Delta S $, as shown in
Fig.~\ref{FIG_S2MOVE}~(a). 

In the similar manner the $ (2,2) $-move as well as the $ (3,1) $-move
can be done by attaching a tetrahedron with 2 or 3 triangles glued to
those neighboring triangles of the $ S^2 $-surface, which can be written in
terms of  $ [\Delta V, \Delta S] $-moves as $ [1,0] $ and $ [1,-2] $
shown in Fig.~\ref{FIG_S2MOVE}~(b) and (c), respectively. 
In addition to these three moves it is also possible to make the three
types of $ (p,q) $-moves by detaching one tetrahedron from 
a $ S^2 $-surface in just the opposite manners to attaching a tetrahedron.... 
It is clear these six moves constitute a set of ergodic moves of
triangulations for the 3-dimensional space with $ D^3 $
topology. 
Each move changes the number of total tetrahedron, $ N_3 $, by one
unit, and at the same time it makes one of the three $ (p,q) $-moves
on the boundary $ S^2 $-surface. 
This method can be extended straightforwardly to other dimensions,
such as the 2-dimensional disk, $ D^2 $, and the 4-dimensional
disk, $ D^4 $ as we will see in the following sections. 
Each case constitutes the four and eight $[\Delta V, \Delta S]$-moves,
respectively.

\section{Creation of simplicial universe}
When the lattice cosmological constant is below the critical value,
the simplicial space tends to expand exponentially as a 
simulation proceeds. 
If we can regard this phenomenon to be the simplicial space analog of what
happened in our universe at its very beginning, the {\it inflation},
the simulation can serve as a good numerical tool to study early stages
 of the evolution of universe. 
However, when we try to simulate the birth of a universe we find
ourselves in a state of complete ignorance: we know neither dynamics
nor initial condition.  
We have to recognize that there is no dynamical law to start with,
since any physical law is considered to appear only at or after the
space-time is settled in its classical sense. 
Also, there is no universe but a fluctuation at the beginning, and
thus no initial condition.
Under these circumstances we are forced to accept {\it a priori} 
elementary rules, but they should be as few as possible 
with the least prejudice. 
We shall employ an analogy of the creation of the natural number space
where two rules known as the Peano axioms are essential: 1) there is an
elementary unit, which is denoted by $ 1 $, and 2) there is the next
neighbor $ S(a) $ of a natural number $ a $.
We are surprised by many profound mathematical laws such as Fermat's
last theorem hidden in this simple space.
\begin{figure}[t]
\begin{center}
\includegraphics[scale=0.7]{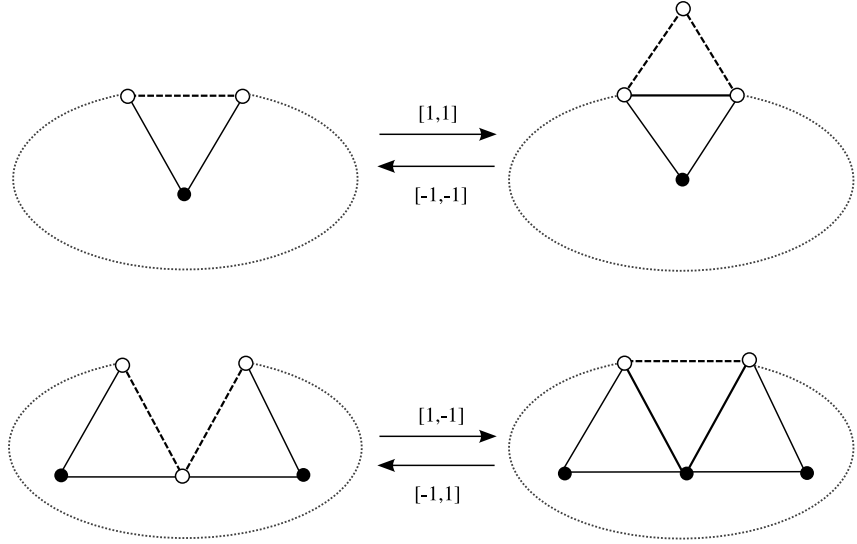}
\end{center}

\caption{2-dimensional $[\Delta V, \Delta S]$-moves. The vertices and
 links on the boundary $S^1$ are denoted as white circles and the dash lines.}
\label{FIG_SVMOVE2D}
\end{figure}

The axioms for the $d$-dimensional simplicial universe we propose are
constituted by 1) there is an elementary unit of universe, which we
choose a $d$-simplex, and 2) there is a set of neighbor universes $
S(\alpha) $, constructed by either attaching or removing an elementary 
$d$-simplex to or from a universe $ \alpha $. 
There will be many neighbor universes possible.
Among them we accept only those universes which fulfill the manifold
conditions: 1) at most two $d$-simplices can attach through one
$(d-1)$-simplex called the {\it face}, and 2) $d$-simplices sharing
one vertex form a $(d+1)$-disk, or a $(d+1)$-semi disk for the case
when the vertex is on the boundary.
According to our least prejudice principle this kind of manifold
conditions may not be necessary, but we expect those universes which
can grow indefinitely large will naturally select configurations
following appropriate manifold conditions from mathematical consistency. 
We shall impose these manifold conditions at the beginning just for
computational simplicity. 
However, the condition may not be unique. 
In Appendix~B we define a different type of moves which create a
manifold having more severely restricted configurations, while the
manifold belongs in the same universality class.
There even exist manifold conditions which allow topology changes.
Inclusion of various possible topologies in the partition function
 is beyond the scope of this paper, and we restrict moves which conserve the 
topology. 

\begin{figure}
\begin{center}
\includegraphics[scale=0.5]{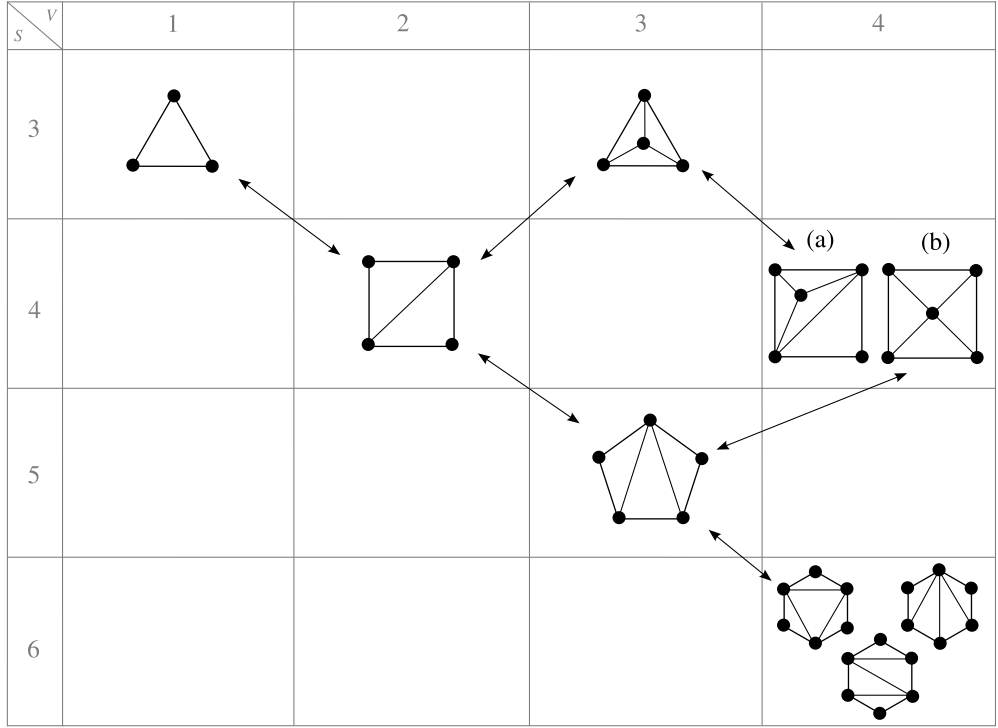}
\end{center}
\caption{2-dimensional universes up to 4 triangles}
\label{FIG_RSNCONFIG4}
\end{figure}
Now, let us start making a two dimensional universe as an exercise.
The 2-dimensional minimum universe is a triangle, which acts also as
the elementary unit.
There exist four moves satisfying the manifold conditions as shown in
Fig.~\ref{FIG_SVMOVE2D}, where we have denoted each move as [$ \Delta
N_2,\Delta \tilde{N}_1 $] with the boundary link number $ \tilde{N}_1 $, 
instead of $ N_1 $, which is related as $ \tilde{N}_1=2N_1-3N_2 $.
Quantization of the simplicial universe is carried out by constructing
the partition function to be a set of all the possible distinct
triangulations. 
Possible configurations are collected as a Markov chain formed by
selecting one of the four moves one after another, with appropriate
weights imposed by the detailed balance condition. 
In Fig.~\ref{FIG_RSNCONFIG4}, we show first few possible
configurations up to universes made of 4-triangles. 
A configuration is specified by four
quantum numbers, ($ N_2 $, $ \tilde{N}_1 $, $ k $, $ s $). 
The third quantum number $ k $ represents the type of a diagram. 
For example, they are assigned to be (a) and (b) in
Fig.~\ref{FIG_RSNCONFIG4} for universes with 
($ N_2 = 4 $, $\tilde{N}_1 = 4$). 
The forth letter $ s $ specifies the point group quantum number of a
diagram with multiplicity $ m_k $, which is $ \tilde{N}_1 $ for most
of the cases except point symmetric diagrams such as
Fig.~\ref{FIG_RSNCONFIG4}~(b) which has $ m_k = 1 $.

\begin{figure}[t]
\vspace*{0.5cm}
\begin{center}
\begin{minipage}{6cm}
\begin{center}
\includegraphics[scale=0.8]{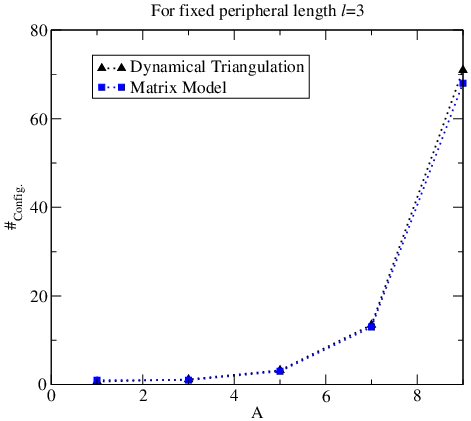}
\caption{Number of configurations for fixed $l$ = 3.}
\label{FIG_FIXL}
\end{center}
\end{minipage}
\hspace{1cm}
\begin{minipage}{6cm}
\begin{center}
\includegraphics[scale=0.8]{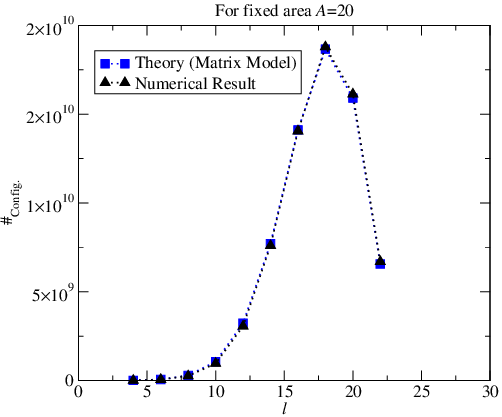}
\caption{Number of configurations for fixed $A$ = 20.}
\label{FIG_FIXA}
\end{center}
\end{minipage}
\end{center}
\end{figure}

In general, moves are independent of the quantum number $ s $, and
when we denote a set of the first three quantum numbers as $ a = \{
N_2, \tilde{N}_1, k \} $, the detailed balance condition reduces to
\begin{equation}
 \frac{p_a m_a}{n_a}w_{a \rightarrow b} = \frac{p_b m_b}{n_b} w_{b
  \rightarrow a}, \label{eq3.1}
\end{equation}
where $ m_a $ is the point group symmetry factor of the configuration $ a $. 
For configurations with large $ N_2 $ there are seldom diagrams having
point symmetries, and we can approximate $ m_a = \tilde{N}_1 $, which
we assume in our simulation. 
This assumption is checked by comparing to the matrix model prediction
(Appendix~A) obtained from expansion coefficients of the generating
function restricting diagrams without tadpoles and self-energy
insertions, for a fixed peripheral length $l$ in Fig.~\ref{FIG_FIXL}
and a fixed area $A$ in Fig.~\ref{FIG_FIXA}.
%

The distribution is obtained by counting the number of states with
equal weight $ p_a = 1 $ for each distinct configuration. 
The equal probability for each distinct configuration is also 
a natural rule for the
quantized universe according to our least prejudice principle.
Besides, we may
 give an arbitral probability $ \{ p_a \} $ through an appropriate
action function $ S_a $ as $p_a = \exp (-S_a)$ depending on the
purpose of a simulation. 
(Here, it may not be relevant to call $S_a$ the action for such a
system without dynamics. 
Nevertheless, we adopt the name only because of the familiarity
accustomed in statistical physics.) 
In $d$-dimension it is written as $
\{N_i\} $, as
\begin{equation}
S^{(d)} \simeq \sum_{i=0}^d \mu^{(i)} N_i.
\end{equation}
By the topological and the manifold constraints, parameters in the
action function will be reduced. 
For the disk topology only the Euler relation is required to satisfy
and the action is expressed as
\begin{equation}
S^{(2)}=\mu N_2 +\mu^B \tilde{N}_1,
 \end{equation}
in 2-dimension, where we use $ \tilde{N}_1 $ instead of the
total number of links $ N_1 $.
We remind these parameters are introduced only for the
convenience to search possible types of universes in the followings.
%
\begin{figure}
\begin{center}
\includegraphics{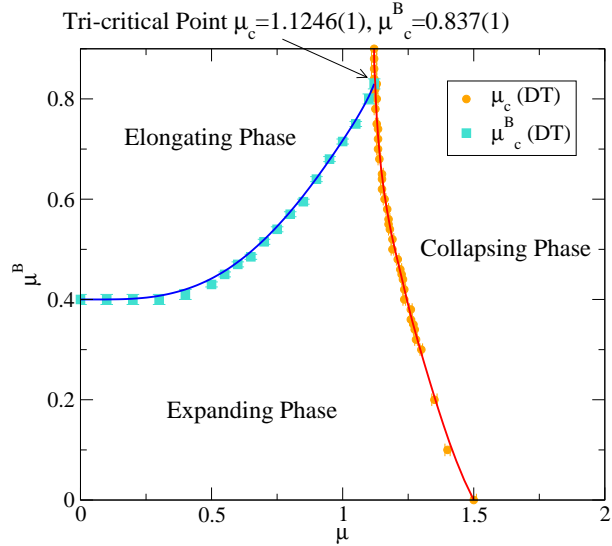}
\end{center}
\caption{Three types of Universes.}
\label{FIG_PHASE2D}
\end{figure}

In the two dimensional parameter space ($ \mu, \mu^B $) the numerical
simulation reveals two critical lines separating into three regions
corresponding to three types of universes as shown in
Fig.~\ref{FIG_PHASE2D}.
When $\mu$ is greater than the critical values $ \mu_c $, the volume
tends to decrease resulting a collapsing universe having a fate to
disappear, while when $ \mu $ and $ \mu^B $ are smaller than the
critical values both the volume and the surface area grow and the
universe keeps expanding. 
The region where $ \mu < \mu_c $ and $ \mu^B > \mu_c^B $ the volume
increases while the surface decreases and the universe looks like an
elongating tube with fluctuating radii. 
The critical parameters $ \mu_c=1.1247(1) $ and $\mu_c^B=0.837(1) $ 
at the tri-critical point coincide to the values predicted by the
matrix model, which we briefly describe in the Appendix A and B for two
types of manifolds.
%

\begin{figure}
\begin{center}
\includegraphics[scale=0.4]{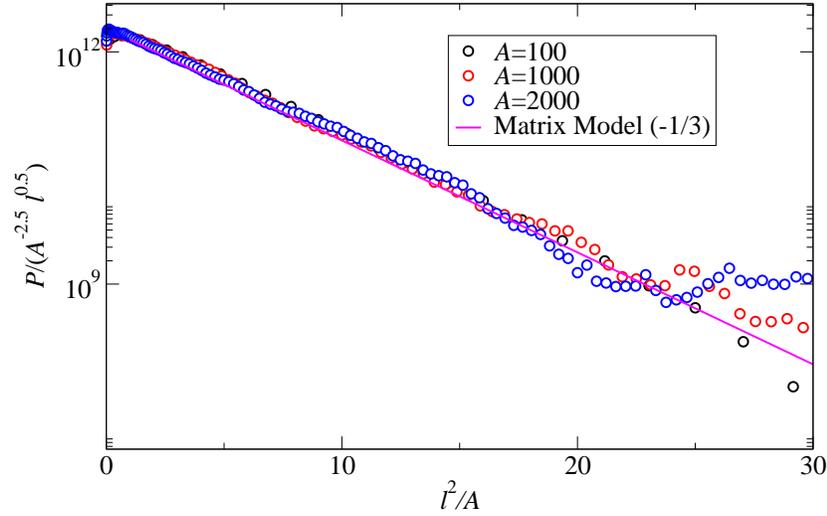}
\caption{The surface area distribution at the tri-critical point for a fixed volume $A=100,1000$, and $2000$.}
\end{center}
\label{FIG_LNP}
\end{figure}
In the numerical simulation, $ \mu_c $ can be easily identified by
inspecting the tendency of volume increase. 
However, the critical point $\mu_c^B $ is not easily determined,
because it does not cause any divergence like $ \mu $.
We fix this parameter so that the surface area distribution for a
fixed volume coincides to the volume-area distribution function
predicted by the matrix model as derived in Appendix~A,
\begin{equation}
P_M (N_2,\tilde{N}_1) \approx N_2^{-2.5} \tilde{N}_1^{0.5} 
 \exp \left(-\frac{1}{3}\frac{\tilde{N}_1^2}{N_2} \right) 
 \exp \{-(\mu-\mu_c) N_2\} \exp\{-(\mu^B-\mu_c^B) \tilde{N}_1\}.
\end{equation}
When we identify the volume and the peripheral length of the boundary Liouville theory and the matrix model as $ A= \frac{\sqrt{3}}{4} a^2 N_2 $ and $ l=a \tilde{N}_1 $, the distribution function (9) is rewritten as 
\begin{equation}
 A^{-2.5} l^{0.5} 
 \exp \left(-\frac{1}{4 \sqrt{3}}\frac{l^2}{A} \right) 
 \exp (-\lambda A) \exp(-\lambda^B l)
\end{equation}
with $ \lambda^B = (\mu^B-\mu_c^B)/a $, which differs from the Liouville theory prediction (C $\cdot$ 10) in two points, namely
the sign in front of the exponent of $ l $, and the factor $ 2 $ in
the coefficient of $\frac{l^2}{A}$. 
These differences is considered to be due to the way of counting the distinct configurations. 
In the boundary Liouville theory those configurations which differ
only in numbering of external lines are thought to be equivalent\cite{AZ}, and the exponent takes minus sign (Appendix~C).
While in the matrix model the point group symmetry factor is taken into
account, and it chooses the plus sign (Appendix~A,and B).
The coefficient of $\frac{l^2}{A}$ depends on the way to choose acceptable diagrams, and in facts, the Tuute algorithm gives $ 192/625 $ instead of $ 1/3 $ as described in Appendix~A,and B.
We can perform simulations either by the matrix model convention or
the Liouville theory convention selecting the multiplicity factor $
m_a $ in Eq.~(\ref{eq3.1}) to be $ {\tilde N}_1 $ or $ 1 $, respectively.
In any way, this ambiguity will not cause any problem practically, since
the multiplicity of a universe cannot be observed. 

The parameter $ b $ appearing in the distribution function is related
to the background charge defined by $ Q=b+b^{-1} $ in the Liouville
theory. 
The exponent of $ A $ is given by 
$ -\left(\frac{Q \chi}{2 b}+\frac{5}{4}\right) $ and for the disk
topology the Eular characteristic $ \chi= 1 $. 
For the pure gravity with the central charge $ c_L=26 $ it is known to
be $ b^2=2/3 $. 
The numerical simulation shows the value very close to the theoretical 
prediction $ -2.5 $. 
The coefficient of $ \frac{\tilde{N}_1^2}{N_2} $ in the exponent near the tri-critical point of the
distribution function appears in the simulation as $ 0.30(4) $ as shown
in Fig.\cite{FIG_LNP} which is consistent
to the theoretical value $ 1/3 $.
From these agreements between simulations and theoretical 
predictions in the distribution function, we feel sure that the 
algorithm can indeed reproduce the universe predicted by the 
2-dimensional quantum gravity.
%

\section{Introducing time and space}
Without knowing what is time the definition is inevitably heuristic 
expecting it will appear naturally reflecting the specific 
property of a universe. 
The reason why there are two types of directions in coordinates, time 
and space, is a reflection of the specific topology of our universe.
For a universe with the disk topology $ D^d $ it is 
natural to identify the space coordinate on the boundary $ S^{d-1} $, 
and the time coordinate along the direction perpendicular to the surface.
A definition of the physical time can be found in the trivial 
relationship between the volume and the area of a universe,
\begin{equation}
 V(t)= \int_0^t S(t') dt', 
\end{equation}
where $ V(t) $ and $ S(t) $ represent the volume ($ \sim N_d a^d $) 
and the surface area ($ \sim \tilde{N}_{d-1} a^{d-1} $), respectively. 
In the numerical simulation for the expanding phase both the volume and
the area increase as the simulation proceeds, which we record as
functions of the {\it diffusion time} $ \tau $, defined by counts of the
number of throwing dice in a Monte Carlo simulation. 
Above relation can be rewritten in terms of $ \tau $ as
\begin{equation}
t=\int_0^\tau \frac{1}{S(\tau')} \frac{d V(\tau')}{d\tau'} d\tau' \label{eq4.1},
\end{equation}
indicating the growing rate of physical time is proportional to the
volume increasing rate per unit area.
The physical time we define has the unique direction of arrow only in
the expanding universe which occurs when the cosmological constant is
below the critical value. 
It becomes negative for the case if the volume decreases in $ \tau $. 

\begin{figure}[t]
\vspace*{0.5cm}
\begin{center}
\includegraphics[scale=0.8]{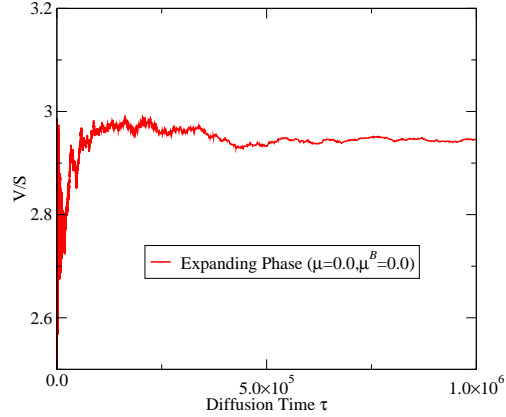}
\caption{The ratio of $V/S$ for the expanding phase.}
\label{FIG_RATIO2D}
\end{center}
\end{figure}
Simulations of the 2-dimensional universe in the expanding phase show the surface area increases
approximately proportional to the volume as
Fig.~\ref{FIG_RATIO2D},
\begin{equation}
 V(\tau) \simeq \sigma^{-1} S(\tau) .
\end{equation}
Thus, from the definition of the physical time, we obtain
\begin{equation}
 t=\sigma^{-1}  \log \left\{ S(\tau)/S(0) \right\},
\end{equation}
or $ S(t)=S(0)e^{\sigma t} $, exhibiting the exponential expansion of the
universe.
As a matter of fact the inflational expansion is a phenomenon we
expect in the classical Liouville theory with the negative
cosmological constant as shown in the Appendix~D.
A homogeneous classical solution for the expanding mode is given by
\begin{equation}
e^{2 b \phi(\eta)}= a^2(\eta) e^{2 b \phi(0)},
\end{equation}
where $a(\eta)=(\cos \omega \eta)^{-1} $ is the conformal scale factor
at the conformal time $ \eta $.
The surface area of a homogeneous universe is the cross-section at $
\eta $ given by $L_\xi /\cos(\omega \eta) $, where $L_\xi$ is the
initial size along the conformal space coordinate $\xi$. 
The volume of the universe at $ \eta $ is then given by 
\begin{equation}
 \int_0^\eta \frac{1}{\cos^2 (\omega \eta')} L_\xi d\eta'.
\end{equation}
We can obtain the physical time through our
heuristic definition, using the volume and the surface area,
 which gives 
\begin{equation}
t=\frac{1}{\omega}\log \left|\frac{1+\sin \omega \eta}{\cos \omega \eta} \right|.
\end{equation}
It coincides with the standard relation between the conformal time and the physical
time, $dt=a(\eta) d\eta$.
Once the time coordinate is determined the space coordinate is defined
on the equal time hyper-surface, $ S^{d-1} $.
In the 2-dimensional universe it is unique to choose the coordinate
axis along the periphery. 

\begin{figure}[t]
\begin{center}
\includegraphics[scale=1]{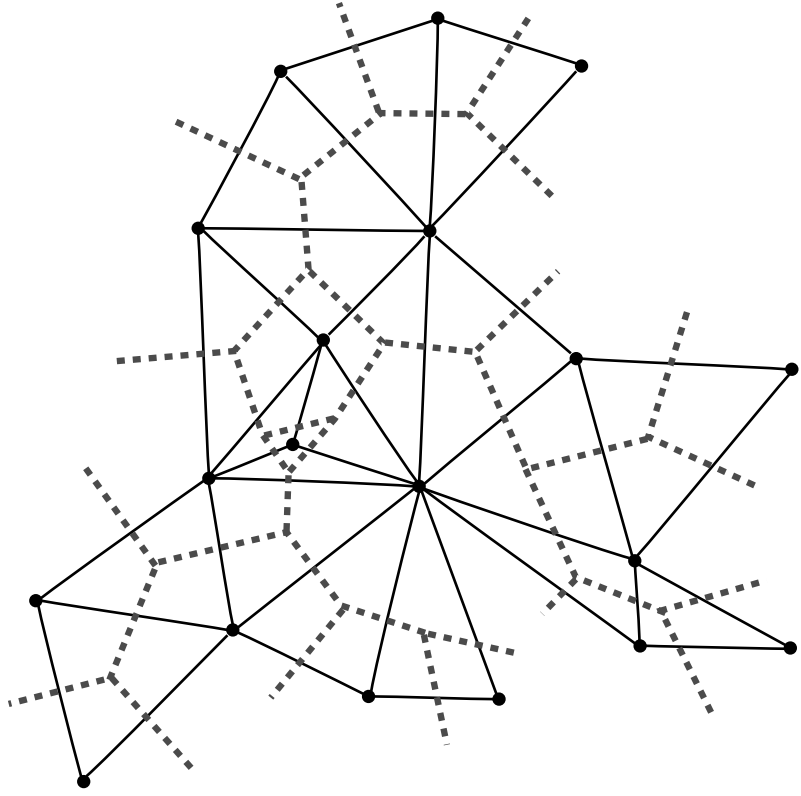}
\end{center}
\caption{Example of the 2-dimensional configuration
 with $(N_2,\tilde{N}_1) = (19,17)$.}
\label{FIG_UNIV2D}
\end{figure}
%
The simplest definition of distance is the geodesic distance counting
the number of links on the hyper-surface connecting two vertices with
the shortest path.
In terms of the distance we can measure the correlation function
between points on the hyper-surface. 
The Liouville theory predicts the two point correlation function of
the boundary primary operator $ B_{\beta}(x) = \exp \{ \beta \phi(x)
\} $ to be 
\begin{equation}
\left<B_{\beta}(0) B_{\beta}(x)\right> \sim \frac {1} {|x|^{2 \Delta_{\beta}}},
\end{equation}
where $ \Delta_{\beta} = \beta (Q- \beta) $.
Assuming the coordination number of a boundary vertex corresponds
 to be the boundary primary operator $ B_{\beta} (x) $ with $ \beta = b $ 
in the simplicial space, we
measured the correlation function of coordination numbers of two
vertices on the boundary using configurations simulated with 
$ \mu = \mu^B = 0 $. 
We found that this correlation function showed a delta-function type
singular behavior which had not been expected at all from the 
Liouville theory. 
\begin{figure}[t]
\begin{center}
\includegraphics{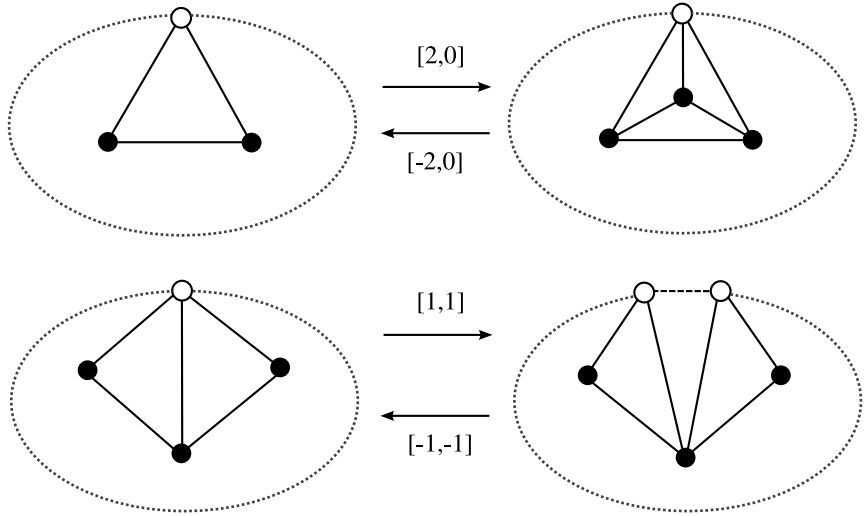}
\end{center}
\caption{Typical $[ \Delta V, \Delta S]$-moves for the Tutte algorithm. The vertices and a link on the boundary $S^1$ are denoted as white circles and a dashed line.}
\label{FIG_SVMOVE2D1PI}
\end{figure}
%
The main reason of this property was thought to come from the singular
nature of the surface created under the moves we have employed. 
The $ [\Delta V,\Delta S] $-moves were designed to generate diagrams
within connected and one-particle irreducible types, forbidding
tad-poles and self-energy insertions at least for the manifolds with
closed topologies. 
However, when we applied it to generate a surface with a boundary, it was
found that the one-particle irreducible nature was lost. 
In fact, most of the configurations had many branches on the boundary
such as the one shown in Fig.~\ref{FIG_UNIV2D} .

The matrix model also contains those diagrams which can be found in
the table for the distribution of diagrams given in Appendix~A.
Under the condition of inflating space, most of the accepted moves are
of the $ [\Delta V,\Delta S]= [1,1] $ type, and it obviously generates an
uncorrelated branches on the boundary.
In order to keep diagrams within the one-particle irreducible type
even for a surface with a boundary, we define a new set of moves as
shown in Fig.\ref{FIG_SVMOVE2D1PI}, consisting $ [\Delta V, \Delta S]
= [{\pm 2},0] $, $ [1,1] $ and $ [-1,-1] $.
The configurations created by these moves coincide with the manifold
studied by Tutte \cite{TUTTE}, which we introduce in Appendix B.
The table for the distribution of diagrams generated by the Tutte
algorithm clearly shows it prohibits branched surface. 
 
\begin{figure}[t]
\begin{center}
\includegraphics{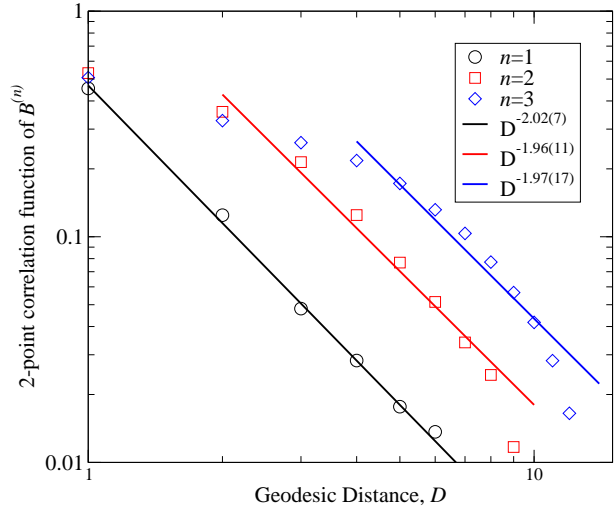}
\end{center}
\caption{The 2-point correlation function for the boundary 
operator $B^{(n)}$.}
\label{FIG_STEPCOR2D}
\end{figure}
%
\begin{figure}
\begin{center}
\begin{minipage}{6cm}
\begin{center}
\includegraphics[scale=0.75]{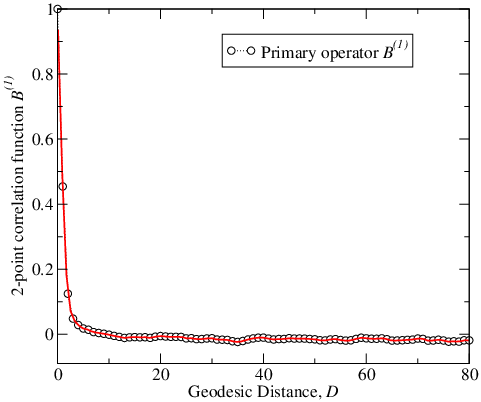}
\end{center}
\caption{2-point correlation averaged over an ensemble of 
$10^4$ universes of typically $N_2=10^3$ sizes.}
\label{FIG_COR2D}
\end{minipage}
\hspace{1cm}
\begin{minipage}{6cm}
\includegraphics[scale=0.7]{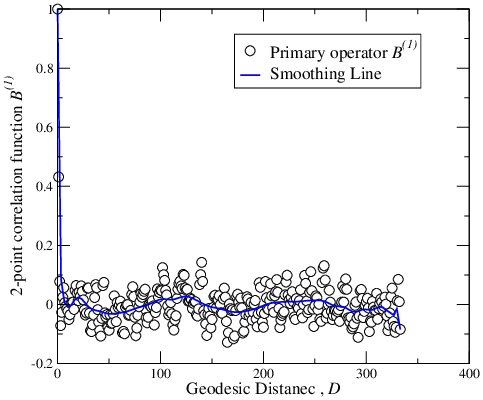}
\caption{2-point correlation function of super-long distances,
 choosing one universe at $\tau=5000$, which has $ N_2 \sim 10^3 $.}
\label{FIG_CORT5000}
\end{minipage}
\end{center}
\end{figure}
%

We measured the correlation function with the new configuration, but
the results still exhibited the singular behavior. 
There seemed yet another important ingredient missing for obtaining
the theoretically expected correlation function. 
We supposed it might be due to the improper definition of the primary
operator in the simplicial space. 
In the previous simulations we chose the local coordination
number as the simplicial primary operator. 
Instead, we defined a new primary operator $ B^{(n)} (i) $ in
the simplicial space by the average of coordination numbers of those
vertices covered within $ n $-steps from a vertex at $ i $.
Results are shown in Fig.~\ref{FIG_STEPCOR2D} for $ n = 1,2,3 $ using a
configuration generated by the Tutte algorithm.
The space dependence shows the predicted exponent, $ -2 $, beyond the
separation $ 2n $ where the overlap of two operators terminates.
We note that the correlation function using the old configuration
remains singular even for the new operator.

In quantum gravity the range of significant correlation is typically the Planck
scale reflecting the characteristic length of dynamical influence.
When the space expands so rapidly, any dynamical perturbation cannot
act on environments, and the quantum fluctuation will be expanded to
macroscopic scale without much disturbance. 
In the construction of the partition function we have assumed that any
possible configuration can be equally a member. 
As the consequence, the two point correlation function averaged over
an ensemble of universes shows the short-distance nature of quantum
mechanical scale shown in Fig.~\ref{FIG_COR2D}, where the long range
correlation specific in each universe is wiped out after averaging
over the ensemble consisting $ 10^4 $ universes of $ N_2 \sim 10^3 $
sizes.
%

\begin{figure}[t]
\begin{center}
\includegraphics[scale=1]{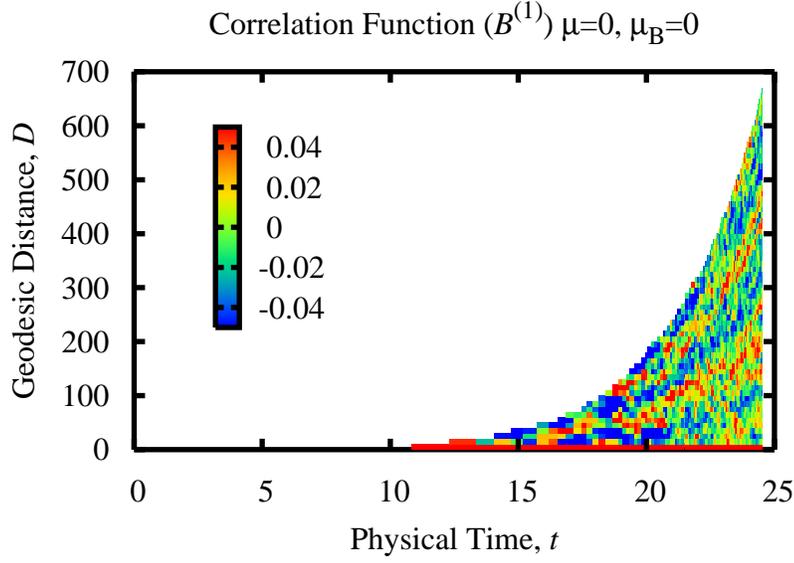}
\caption{Evolution of the 2-point function in the physical time $t$.}
\label{FIG_CORTRACE}

\end{center}
\end{figure}
\begin{figure}
\begin{center}
\vspace{1cm}
\includegraphics[scale=0.8]{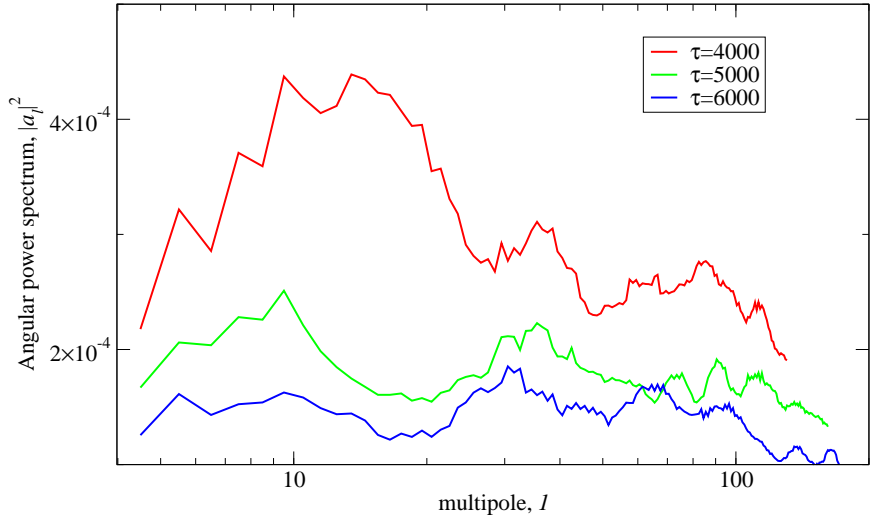}
\caption{Power spectrum at $\tau=4000, 5000 \, {\rm and}\,  6000$.}
\label{FIG_SPT5000}
\end{center}
\end{figure}
%
In this respect one of the most interesting phenomena of the
space-time geometry can be found in the observations of CMB
anisotropies. 
An ultimate source of the anisotropies only we can imagine is
the accidental creation of a seed universe by quantum fluctuation, and
random expansion of space by accumulations of elementary units
according to our model. 
When we measure the two-point correlation function, selecting one
universe as COBE and WMAP observations did, it exhibits significant
super-long distance correlation as shown in Fig.~\ref{FIG_CORT5000}.
The pattern of this fluctuation seems to be conserved in some degree
during the expansion (Fig.~\ref{FIG_CORTRACE}) as we have imagined in
analogy to a picture on an inflating balloon. 
We also calculate the angular power spectrum of the two point
correlation function defined by $ | a_l |^2 $ with
\begin{equation}
a_l=\int d \theta P_l (\cos \theta) f( \cos \theta),
\end{equation}
where $ \theta = x/r(\tau) $ is the ratio of geodesic distance and the
peripheral length at $ \tau $.
It shows the persistence of long distance correlation 
Fig.~\ref{FIG_SPT5000}.
%
%

\section{Discussions and Conclusion}
Stimulated by the recent observations on the CMB anisotropies, a model
to construct a quantized universe is proposed. 
Reexamining the DT-method developed in the simplicial quantum gravity,
we extend the algorithm so that it can create a manifold with a
boundary.
Numerical simulation is made mostly on the 2-dimensional universe. 
Three types of universes, expanding, elongating and collapsing, are
observed by varying two cosmological constants.
Numerical results are closely examined by comparing to the matrix
model and the Liouville theory predictions, which are considered to be
the consistent theories of quantum gravity in 2-dimension. 
The results exhibits fair agreements to the theoretical predictions. 
Physical time is defined based on the standard relation between volume
 and surface area, and the surface expands exponentially 
in the physical time.
The exponential expansion is regarded as the quantum
gravity realization of the inflation of the universe.
Then, the two
point correlation function on the equal-time hyper-surface is
analyzed as the function of the geodesic distance. 
It shows the proper exponent expected by the Liouville field theory
 in the quantum scale after averaging over an ensemble of universes,
 while each universe has an extremely long distance correlation.
The anisotropies originated from quantum
fluctuations and the subsequent inflationary expansion is one
of the most significant findings of COBE and WMAP missions.

Among many problems we need to clarify, the next step we should take
is a numerical study on the universe in 4-dimension.
We have already reported a few preliminary works on the algorithm and
numerical results in several occasions \cite{qc4}. 
In 4-dimension there are eight [$ \Delta V, \Delta S $]-moves,
corresponding to the four ($ p,q $)-moves of the 3-dimensional DT
and two types of attaching or removing a 4-simplex to or from the $
S^3 $ boundary for each ($ p,q $)-move. 
Starting with a 4-simplex the Monte Carlo simulation is carried out
for the partition function by constructing a Markov chain under the
detailed balance condition accepting moves which fulfill the manifold
conditions in exactly the same manner as the 2-dimensional
simulation. 
The action to be used for the a priori probability $ p_a \simeq
\exp(-S_a) $ is assumed to be
\begin{equation}
 S=\mu N_4+\mu^B \tilde N_3+\kappa N_2+\kappa^B \tilde N_1,
 \end{equation}
where $ \kappa $ and $ \kappa^B $ are referred to be the gravitational
constant and the boundary gravitational constant. 
Extra two boundary terms are due to lack of relations existing in
closed manifolds. 
Varying these parameters we can obtain several types of universes.
For example, three kinds of universes similar to the 2-dimensional
case are observed by varying $ \mu $ and $ \mu^B $ while two
parameters, $ \kappa $ and $ \kappa^B $, are fixed to be zero.
The expansion is exponential type in terms of the physical time 
in the simulation with $ \mu = \mu^B =0 $.

\begin{figure}
\begin{center}
\includegraphics[scale=1]{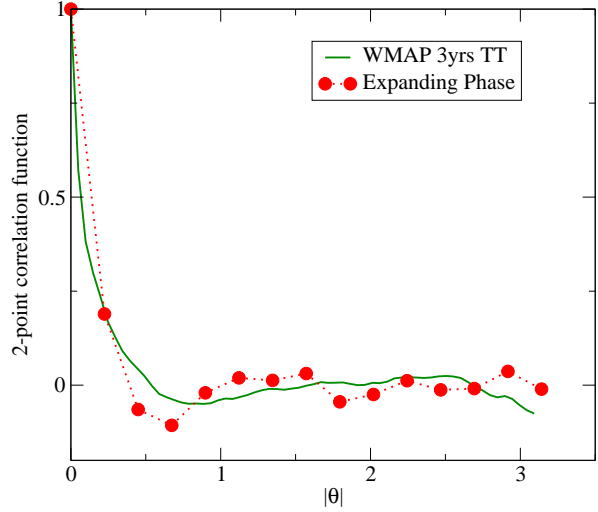}
\end{center}
\caption{Comparison of 2-point correlation functions of the
 4-dimensional simulation and the WMAP third year data. Normalization
 constant is $f(0)$ for both cases.}
\label{FIG_COR4D}
\end{figure}

The equal-time hyper surface $ S^3 $ is considered to be a model for 
the very early universe.
In order to appeal the fair possibility of the method we present 
here a comparison of the two point correlation function obtained
from simulations and the WMAP data.
The operator is chosen to be the scalar curvatures which are
defined by the number of 4-simplices sharing one triangle.
We use the simulation data of the space with the topology $ D^4 $
and $ N_4 \sim 10,000 $. 
In its $ S^3 $ boundary we create a $ B^3 $ cavity removing about half
of the total 4-simplices covered within certain geodesic distance from
an arbitrarily selected 4-simplex, and regard the $ S^2 $ boundary with the area $ \tilde{N}_2 $ as the last scattering surface(lss). 
Then, we choose one triangle $ i $ on the lss and collect all the
triangles $ \{ j(i,x) \} $ with geodesic distance $ x $ from $ i $ measured in the dual lattice space.
The two point correlation function is defined by
\begin{equation}
 f(x) = \frac{1}{\tilde{N}_2}
        \sum_{i=1}^{\tilde{N}_2}
          \frac{1}{\tilde{n}_2(i,x)} 
            \sum_{j(i,x)} 
            \frac{(R_i-\bar{R})}{\bar{R}} 
            \frac{(R_j-\bar{R})}{\bar{R}},
 \end{equation}
where $ R_i $ and $ \bar{R} $ are the scalar curvature at a triangle 
$ i $ and the average over $ \tilde{N}_2 $ triangles, respectively.
In order to compare to the observations, we also measure the
two point correlation function of WMAP\cite{WMAP-DATA} in the same manner as we have done for the simulation data: 
choose a point $ i $, collect points $ \{ j(i,x) \} $ with the distance
 $ x $ apart from $ i $, and calculate the temperature correlation function.
The results are compared in Fig.~\ref{FIG_COR4D}.
As far as the two point correlation function is concerned the agreement to the
 observation data is significant comparing to the $ \Lambda $CDM\cite{WMAP}. 

Studies on the effects of extra two parameters to the structure of
universe is one of the imminent problems.
As in the 2-dimensional case, there is no reason to select specific
values for these parameters.
If the nature selects any, it should be the consequence of additional
degrees of freedom.
Therefore, the most important step we have to take is to go beyond the
pure gravity by introducing the matter degrees of freedom. 
Without matter fields only the inflational expansion will continue
forever. 
The universe is considered to have made a transition from the
inflation to the big-bang by the deceleration effect of matter
fields\cite{HHY-PR}.
Following the same kind of procedures as the construction of
simplicial manifold the additional variables are attached to each
sub-simplex with appropriate rules for defining the neighbors. 
We should remember at this moment there exists no {\it a priori} laws
of physics and thus no action function.
The laws of physics which we expected to appear from the appropriate
rules are the relativistic kinematics and quantum dynamics, which
will be a primitive form of the relativistic quantum field theory.

We have just started the study of the origin of a universe in 2-dimension 
as the first step. It is a small universe, but the necessary step toward 
the universe in 4-dimension. 
After all there should be universes in any dimension, if it is
mathematically consistent.
We recall the a mathematician's monologue\cite{HARDY} `Is mathematics
reality?'.
We also agree his answer the reality lies outside us, and we physicists discover dynamical laws, which is a reflection of the mathematical reality....

We would like to give our sincere thanks to Dr.'s K.~Hamada,
A.~Zamolodchikov, and H.~Kawai for their stimulating discussions
 and especially N.~D.~Hari Dass for his careful reading of the manuscript....

\appendix
\section{Matrix model}
\begin{figure}[b]
\begin{center}
\includegraphics[scale=0.6]{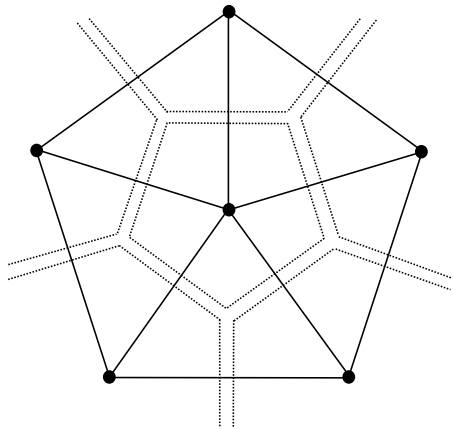}
\caption{Correspondence diagram between the matrix model and the triangulated
manifold}
\end{center}
\label{FIG_MAT}
\end{figure}
The one matrix model of cubic vertex is defined by the partition function,
\begin{equation}
 Z_M=\int D[M] \exp
  \left\{ -\left(\frac{1}{2}{\rm tr} M^2 - g {\rm tr} M^3\right) \right\},
\end{equation}
with $ N \times N $ Hermitian matrix $ M $ as the dynamical variable, where
\begin{equation}
 D[M]= \prod_i dM_{ii} \prod_{i>j} d[\Re M_{ij}] d[\Im M_{ij}] . 
\end{equation}
Our main interest is one point Green's function,
\begin{equation}
 G_l(g) = \int D[M] {\rm tr} (M^l) \exp 
 \left\{ -\left(\frac{1}{2}{\rm tr} M^2 - g {\rm tr} M^3\right) \right\},
\end{equation}
and its generating function $ F(j,g)=\sum_{l} G_{l}(g) j^l $. 
Expanding $ F(j,g) $ in powers of $ g $ and $ j $ as $ \sum_{l,k}
f_{l,k} g^k j^l $ the coefficient $ f_{l,k} $ represents the number of
diagrams with $ k $ vertices and $ l $ external lines. 
The significance of this coefficient lies on the fact that the dual
graph of a diagram corresponds to the two dimensional triangulated
manifold with $ k $ triangles and $ l $ boundary links as Fig.~A$\cdot$1.

After a few steps of mathematical manipulations for selecting
connected planer diagrams excluding tadpoles and self-energy
insertions, the equation for the generating function is obtained to
be\cite{BIPZ}
\begin{equation}
g F^2 - [g + j (1+s) ]F +j(1+s) -g j^2 +j^3 = 0
\end{equation}
with  $ s(g) =g G_3(g) $. 
The first few components of Green's function, $ G_0(g)=1, G_1(g)=0 $
and, $ G_2(g)=1 $, reflect the constraints which we have imposed as
the acceptable diagrams. 
This equation contains two unknown functions, $ F(j,g) $ and $ s(g) $. 
By requiring that they are both analytic at $ g=j=0 $, $ s(g) $ is
found to satisfy
\begin{equation}
s^4+3 s^3+8 g^2 s^2 +3 s^2-20 g^2 s+s+16 g^4-g^2= 0.
\end{equation}
Expanding these equations in terms of $ g $ and $ j $ recursively we
obtain the distribution $ f_{l,k} $. 
We list first several coefficients in Table ~\ref{TABLE_BIPZ3}. 
From this table we notice that the branching polymer type diagrams
such as the one with $ k $ vertices and $ k+2 $ external lines are not
forbidden.
Diagrams we have accepted are the one particle irreducible type only
for those without boundaries or a minimum boundary with 3-external
lines. 
In the subsequent Appendix we introduce another type of generating
function which counts only diagrams of one particle irreducible type
even with a boundary.
\begin{table}[t]
\caption{The coefficients $ f_{l,k} $ -the distribution of diagrams with $ l $-links and $ k $-triangles.}
\label{TABLE_BIPZ3}
\begin{tabular}{c|rrrrrrrrrr}
$(l|k)$ & 1 & 2 & 3 & 4 & 5  & 6  & 7 & 8 & 9 & 10 \\ \hline
     3  & 1 & 0 & 1 & 0 & 3  & 0  &  13  & 0 & 68 & 0  \\
     4  &   & 2 & 0 & 5 & 0  & 20 & 0   & 100 & 0 & 570  \\
     5  &   &   & 5 & 0 & 21 & 0 & 105 & 0 & 595 & 0  \\
     6  &   &   &   & 14 & 0 & 84 & 0 & 504 & 0 & 3192  \\
     7  &   &   &   &   & 42 & 0 & 330 & 0 & 2130 & 0  \\
     8  &   &   &   &   &   & 132 & 0 & 1287 & 0 & 10296  \\
     9  &   &   &   &   &   &   & 429 & 0 & 5005 & 0  \\
    10  &   &   &   &   &   &   &   & 1430 & 0 & 19448  \\
    11  &   &   &   &   &   &   &   &   & 4862 & 0  \\
    12  &   &   &   &   &   &   &   &   &   & 16796  \\
\end{tabular}
\end{table}

From the singularity analysis, $ s(g) $ is shown to be singular at $
g=g_c=\sqrt{27/256} $, and $ F(j,g) $ becomes singular at  $ g=g_c $,
$ j=j_c=\sqrt{3}/4 $. 
Logarithm of these two critical couplings correspond to the critical
(lattice) cosmological constant $ \mu_c = 1.125 $ and the critical
(lattice) boundary cosmological constant $ \mu_c^B = 0.837 $.
Expanding at the tri-critical point the first three leading singular terms of the generating function is given by
\begin{equation}
F(j,g) \sim (j-j_c)^{-\frac{3}{2}} (g-g_c)^{\frac{3}{2}}+2^{-1} 3^{-\frac{1}{2}} (j-j_c)^{-\frac{7}{2}} (g-g_c)^{\frac{5}{2}}+2^{-3} (j-j_c)^{-\frac{11}{2}}(g-g_c)^{\frac{7}{2}}+ \dots .
\end{equation}
Using the binomial expansion formula, $ (x-x_c)^{\alpha} = \sum_n
 Z^{\alpha}_n (x_c)^n $, with $ Z^{\alpha}_n \sim \frac{\sin\pi \alpha}{\pi} (-1)^{1+\alpha}  n^{-(1+\alpha)} \Gamma (1+\alpha)
 (g_c)^n $ for a large $ n $, the expansion coefficient of the distribution function can be written as 
\begin{equation}
f_{n,m} \sim m^{-2.5} n^{0.5} \exp\{-(\mu-\mu_c) m\} \exp\{-(\mu^B-\mu_c^B) n\} (1 - \frac{1}{3} \frac{n^2}{m} + \frac{1}{18} \frac{n^4}{m^2} +\dots) .
\end{equation} 
If we assume the exponential form as eq.(9) of the Liouville theory prediction the distribution function of the matrix model is expected to be
\begin{equation}
f_{n,m} \sim m^{-2.5} n^{+0.5} \exp\{-(\mu-\mu_c) m\} \exp\{-(\mu^B-\mu_c^B) n \} \exp\left(-\frac{1}{3} \frac{n^2}{m} \right) ,
\end{equation} 
or in terms of the area $ A= \frac{ \sqrt{3}}{4} a^2 m$ and the peripheral length $ l=a n $, it is written as
\begin{equation}
P_M (l,A) \sim A^{-2.5} l^{0.5} \exp(-\lambda A) \exp(-\lambda^B l) \exp\left(-\frac{1}{4 \sqrt{3}} \frac{l^2}{A}\right) .
\end{equation} 
Difference of the coefficient of the $ \frac{l^2}{A}$ from the Liouville theory by the factor $ 2 $ is also due to the symmetric factor.
\section{Tutte algorithm}
Tutte has been successful to obtain an equation for the generating
function of the distribution of diagrams in the cubic vertex matrix
model restricting within connected, and one-particle irreducible
types. 
The generating function $ T(y,x) $ are expanded as  $ \sum_{m,n}
t_{m,n} x^n y^m $, where $ n $ is the number of loops and $ m+3 $ is
the number of external lines. 
The exponent $ m $ and $ n $ are related to the number of vertices $ k
$ and links $ l $ of the last Appendix as $ k=2n+m+1 $ and $ l=m+3
$.
The equation for $ T(y,x) $ satisfies,

\begin{equation}
y^2 T^2 + (x+x{\tilde s} y-y-y^2) T + y-x {\tilde s}=0,
\end{equation}
where the unknown function $ {\tilde s}(x) = T(0,x) $ is determined by
the analyticity condition of $ T(y,x) $ at $ x=y=0 $, which leads
\begin{equation}
(8x^2{\tilde s}^2+-20x{\tilde s}+32x-1)^2+(8x{\tilde s}-1)^3=0.
\end{equation}
This equation is the same as Eq($ A \cdot 4 $), if we set $ s =
x{\tilde s} $ and $ x= g^2 $.
The importance of the Tutte distribution function is the one-particle
irreducible property of diagrams with a boundary. 
For example, the branching polymer type diagrams of $ m+3 $ external
lines with zero loops are forbidden as $ t_{m,0}=\delta_{m,0} $.
The first few terms of the coefficients are listed in the
Table~\ref{TABLE_TUTTE}.
 \begin{table}[t]
\caption{The coefficients $ t_{m,n} $ -the distribution of diagrams with $ (m+3) $-links and $ n $-loops.}
\label{TABLE_TUTTE}
\begin{tabular}{c|rrrrrrrrrr}
$(m|n)$ & 0 & 1 & 2 & 3 & 4  & 5  & 6 & 7 & 8 & 9 \\ \hline
     0  & 1 & 0 & 1 & 0 & 3  & 0  &  13  & 0 & 68 & 0  \\
     1  &   & 0 & 0 & 1 & 0 & 6 & 0  & 36 & 0   & 228   \\
     2  &   &   & 0 & 0 & 1 & 0 & 10 & 0 & 80 & 0   \\
     3  &   &   &   & 0 & 0 & 1 & 0 & 15 & 0 & 155   \\
     4  &   &   &   &   & 0 & 0& 1 & 0 & 21 & 0   \\
     5  &   &   &   &   &   & 0 & 0& 1 & 0 & 28   \\
     6  &   &   &   &   &   &   & 0 & 0& 1 & 0   \\
     7  &   &   &   &   &   &   &   & 0 & 0& 1  \\
     8  &   &   &   &   &   &   &   &   & 0 & 0  \\
     9  &   &   &   &   &   &   &   &   &   & 0   \\
\end{tabular}
\end{table}

The singularity of $ {\tilde s}(x) $ is identical to that of $ s(x) $
of Appendix~A at $ x=x_c= 27/256 $, and $ T(y,x) $ becomes singular
at $ x=x_c $,  $ y=y_c=3/16 $, or in terms of the variables $ g $ and
$ j $ it is at $ g_c = \sqrt{x_c} $, $ j_c = \frac{y_c} {\sqrt{x_c}}$....
Logarithm of the surface critical couplings correspond to the critical
(lattice) boundary cosmological constant $ \mu_c^B = 0.5493 $, which
differs from the value given in Appendix~A.
The generating function is expanded near the singularity similar to the last Appendix as 
\begin{equation}
T(y,x) \sim (j-j_c)^{-\frac{3}{2}} (g-g_c)^{\frac{3}{2}}+3^{\frac{3}{2}} (\frac{2}{5})^{5} (j-j_c)^{-\frac{7}{2}} (g-g_c)^{\frac{5}{2}}+ 3^4 (\frac{2}{5})^9 (j-j_c)^{-\frac{11}{2}} (g-g_c)^{\frac{7}{2}}, 
\end{equation}
which lead the expansion coefficient of the distribution function to be 
\begin{equation}
f_{n,m} \sim m^{-2.5} n^{+0.5} \exp\{-(\mu-\mu_c) m \} \exp\{-(\mu^B-\mu_c^B) n\} \exp(-\frac{192}{625} \frac{n^2}{m}) .
\end{equation} 
The exponents coincide with the previous case, and the distribution
function has the same asymptotic form as the previous case with the
different value for $ \mu_c^B $ and the coefficient in front of $ \frac{n^2}{m} $.

We also can construct the moves which generate the diagrams satisfying
the Tutte manifold conditions. 
They constitute four moves: $ [\Delta V, \Delta S] = [2,0]$, and $
[1,1]$, and their inverse moves, $ [-2,0] $, and $ [-1,-1] $. 
They are shown graphically in Fig.~\ref{FIG_SVMOVE2D}.

\section{Boundary Liouville field theory}
The partition function of the Liouville field theory with the
conformally invariant boundary is defined by\cite{FZZ}
\begin{equation}
Z_L= \int D[\phi] \exp\left(-A_{bulk} - A_{bound}\right).
\end{equation}
The bulk Liouville action $ A_{bulk} $ is given by
\begin{equation}
A_{bulk}=\frac{1} {4\pi} \int_\Gamma 
\left[\hat{g}^{ab} \partial_a\partial_b \phi + Q \hat{R} \phi + 4 \pi \lambda e^{2 b \phi}\right] \sqrt{\hat{g}} d^2x,
\end{equation}
where $ \hat{R} $ is the scalar curvature associate with the
background metric $ \hat{g}_{ab} $ while $ Q(=b+1/b) $ is the
background charge. 
We consider only the geometry of a disk which can be represented as a
 simply connected domain $\Gamma $.
The conformally invariant boundary action is written as
\begin{equation}
A_{bound} =  \frac{1}{ 2 \pi} \int_{\partial \Gamma} 
\left[Q \hat{K} \phi + 2 \pi \lambda^B e^{b \phi}\right] \hat{g}^{1/4} d\xi,
\end{equation}
where $ \hat{K} $ is the scalar curvature of the boundary associated
with the background metric $ \hat{g}_{ab} $. 
The integration over $ \xi $ is taken along the boundary. 

Since the partition function is invariant under a constant shift 
$\phi \rightarrow \phi +  \frac{\sigma}{2b} $, the scaling relation,
\begin{equation}
Z[\lambda,\lambda^B]=e^{- \frac{\sigma}{2b} \chi Q} Z[\lambda e^\sigma,\lambda^B e^{\sigma /2}],
\end{equation}
holds, where $ \chi $ is the Euler characteristic of the 2-dimensional
surface. 
Choosing $ e^\sigma = \lambda^{-1} $, the partition function is expected
to be the homogeneous function,
\begin{equation}
 Z[\lambda,\lambda^B]=\lambda^{\chi Q/2b} Z[1, \frac{\lambda^B}{\sqrt{\lambda}}].
\end{equation}
This relation can be written in terms of the Laplace transformed variables defined by
\begin{equation}
 Z[\lambda,\lambda^B]=\int_0^\infty dA \int_0^\infty dl~ e^{-\lambda A - \lambda^B l} \tilde{Z}[A,l],
\end{equation}
where $ A $ and $ l $ are the variables corresponding to the area and
the boundary length of the 2-dimensional manifold, respectively.
In terms of these variables the scaling relation reads,
\begin{equation}
 \tilde{Z}[A,l]=e^{-\frac{\sigma}{2b} \chi Q-\frac{3 \sigma}{2}} \tilde{Z}[e^{-\sigma} A, e^{-\sigma/2} l].
\end{equation}
Choosing $ e^\sigma =A $, we have
\begin{equation}
\tilde{Z}[A,l]=A^{-\left( \frac{\chi Q}{2b} + \frac{3}{2}\right)}\tilde{Z}[1,\frac{l}{\sqrt{A}}].
 \end{equation}
According to Ref\cite{FZZ} the scaling function is given by,
\begin{equation}
 \tilde{Z}[1, \frac{l}{\sqrt{A}}]=
 \left(\frac{l}{\sqrt{A}}\right)^{\frac{Q}{b}-3} 
 e^{-\frac{1}{4 \sin \pi b^2}\frac{l^2}{A}}.
 \end{equation}
For the case of pure gravity, $ b=\sqrt{2/3} $, with the disk
topology, $ \chi=1$, the distribution function reads,
\begin{equation}
\tilde{Z}[A,l]=A^{ -2.5}l^{-0.5}
 \exp \left( -\frac{1}{2 \sqrt{3}} \frac{l^2}{A} \right).
 \end{equation}
which differs from the matrix model prediction by a factor $ l $. 
The partition function is given by
\begin{equation}
Z[\lambda,\lambda^B]= \lambda^{5/4} 
\left( 1+\sqrt{\sin \pi b^2} \frac{\lambda^B}{\sqrt{\lambda}} \right)^{3/2} 
\left( 1-\frac{2}{3} \sqrt{\sin \pi b^2} \frac{\lambda^B}{\sqrt{\lambda}}\right),
\end{equation}
apart from a numerical factor.

There are various correlation functions known analytically. 
Among them we write here necessary correlation functions for the
primary vertex operators, $V_b = e^{2 b \phi} $ and $B_b =  e^{b \phi}
$,which correspond to the bulk density operator and the boundary
density operator, respectively, 
\begin{eqnarray}
 \left<V_b(z)\right> &\simeq& \frac{U(b)}{|z-\bar{z}|^{2 \Delta_b}} \\
 \left<B_b(0) B_b(x)\right> &\simeq& \frac{d(b)}{|x|^{2 \Delta_b}},
\end{eqnarray}
where the dimension $ \Delta_{\beta}=1 $ for $ \beta = b $.
The coefficients $ U(b) $ is related to the distribution function as $
U(b) \sim - \frac{\partial}{\partial \lambda} Z[ \lambda,\lambda^B] $, 
\begin{equation}
U(b) \sim \lambda^{1/4} 
\left( 1+\sqrt{\sin \pi b^2} \frac{\lambda^B}{\sqrt{\lambda}}\right)^{1/2} ,
\end{equation}
while the coefficient $ d(b) $ is proportional to $
\frac{\partial^2}{\partial (\lambda^B)^2} Z[ \lambda,\lambda^B] $, 
\begin{equation}
d(b) \sim \lambda^{1/4} 
\left( 1+\sqrt{\sin \pi b^2} \frac{\lambda^B}{\sqrt{\lambda}}\right)^{-1/2}
\left( 1+2 \sqrt{\sin \pi b^2} \frac{\lambda^B}{\sqrt{\lambda}}\right) ,
\end{equation}
within a numerical factor.

\section{Classical Liouville equation}
The Liouville field theory is considered to be valid only in the
region where the stable vacuum exists. 
However, solutions of the classical equation of motion derived
from the Liouville action,
\begin{equation}
 S_L = \frac{1}{4 \pi}\int d\eta d\xi \left[ (\partial_\eta \phi)^2
  -(\partial_\xi \phi)^2+4 \pi \lambda e^{2 b \phi}\right], 
\end{equation}
are known after J.~Liouville.
Here, the metric is taken to be Lorenzian and the sign of the cosmological
constant is opposite to the Liouville field theory. 
The classical equation is written as,
\begin{equation}
 \partial_{\eta}^2 \phi = \partial_{\xi}^2 \phi + 4 \pi \lambda b e^{2 b \phi}
\end{equation}
Changing the variables as $ y=\xi+\eta $ and $ z=\xi-\eta $, the
solution $ w(y,z) $ follows the equation
\begin{equation}
\partial_y \partial_z w = - \pi \lambda b e^{2 b w}.
\end{equation}
The solution is known to have the following form, 
\begin{equation}
e^{2 b w}= - \frac{1}{\pi \lambda b^2} \frac{(\partial_z A)(\partial_y B)}{\{ 1-A(z) B(y) \}^2 } ,
\end{equation}
where unknown functions $ A(z) $ and $ B(y) $ are fixed by the initial
conditions. 

We consider a case when the initial conditions are given by
\begin{equation}
e^{2 b \phi(0,\xi)} = \rho_0(\xi), ~\partial_\eta \phi(\eta,\xi)|_{\eta=0} =0.
\end{equation}
The second condition requires $ B(\xi)= \pm A(\xi) $. From the
positivity of the density, $ e^{2 b \phi(\eta,\xi)} $, we need to
choose the minus sign for $ \mu>0 $ case. 
Writing $ \rho(\eta,\xi) = e^{2 b \phi(\eta,\xi)} $ the solution is
given by
\begin{equation}
 \rho(\eta,\xi) = \frac{\rho_0^{1/2} (\xi-\eta) \rho_0^{1/2}{(\xi+\eta)}}{{\cos^2 \{ F(\xi-\eta)-F(\xi+\eta) \} }},
\end{equation}
where
\begin{equation}
 F(\xi) = (\pi \mu b^2)^{1/2} \int^\xi \rho_0^{1/2} (\xi')d\xi'.
\end{equation}
For the homogeneous case, $ \rho_0(\xi) = \rho_0 $ (Const.), the
solution is given by
\begin{equation}
 \rho(\eta)=\frac{\rho_0}{\cos^2(\omega \eta)} ,
  ~\omega=(4 \pi \lambda b^2 \rho_0)^{1/2}.
\end{equation} 
Then the conformal scale factor $ a(\eta) = e^{b \phi} $ is written as
$a(\eta)=\cos^{-1}(\omega \eta )$, and the physical time $ t $ defined
by $ dt = a(\eta)d\eta $ is expressed as
\begin{equation}
 t= \frac{1}{\omega} 
\log\left|\frac{1+\sin\omega\eta} { \cos \omega\eta }\right|,
\end{equation} 
or $ \cosh \omega t = 1/ \cos (\omega \eta) $. 
The boundary length $ a(\eta) l_0 $ represented in terms of the
physical time is expanding exponentially as $ l(t) = l(0) \cosh(\omega
t) $.


\end{document}